%% file: main.tex
\documentclass[%
 twocolumn,
 amsmath,amssymb,
 aps,
 floatfix,
]{revtex4-2}

\usepackage{amssymb,amsmath,amstext}%%   American Physical Society math etc extensions
\usepackage{graphicx}% Include figure files
\usepackage{bm}% bold math
\usepackage{hyperref}
\let\olddiv\div
\usepackage{physics}
\usepackage{ulem}
\usepackage{enumitem}
\usepackage[OT4]{fontenc}
\usepackage{lipsum}

\DeclareMathOperator{\sign}{sign}

\begin{document}

\title{Interaction-driven losses for atoms in a dark-state lattice}

\author{Piotr Kubala}
\affiliation{Institute of Theoretical Physics, Jagiellonian University in Krak\'ow, \L{}ojasiewicza 11, 30-348 Krak\'ow, Poland }
\author{Mateusz \L\k{a}cki}
\affiliation{Institute of Theoretical Physics, Jagiellonian University in Krak\'ow, \L{}ojasiewicza 11, 30-348 Krak\'ow, Poland }

\begin{abstract}
In this work we estimate the collisional loss rate of ultracold bosons in the optical potential featuring subwavelength-width peaks. This is established by using $\Lambda$ arrangement of three atomic states coupled (almost) resonantly by lasers. Using Fermi's Golden Rule, we find that the loss rate is influenced by the overall strength of the lasers, with the largest losses occurring when the two-photon transition is blue-detuned from the excited state of the $\Lambda$ system. Overall, the predicted loss rates are low, which may allow the use of ultracold bosons in the construction of dark-state potentials in the $\Lambda$-type many-level system.
\end{abstract}

\maketitle

\section{Introduction}

The optical lattice potentials are periodic potentials for ultracold atoms usually created via an AC-Stark shift in a
two-level far-detuned system \cite{Bloch2002}. They allow one to experimentally implement Hamiltonians
important for condensed matter physics \cite{bloch2008many,lewenstein2012ultracold,bloch2012quantum,dutta2015} and tune system parameters including the amplitude for
atom hopping and strength of inter-atom interactions \cite{fedichev1996influence,bohn1997prospects,theis2004tuning,inouye1998observation,chin2010feshbach,ciurylo2005optical}.

When gas populates only the lowest Bloch band of a deep optical lattice, it is modelled well by a tight-binding description -- a Bose-Hubbard model \cite{jaksch1998cold}. It includes a term proportional to $U$ describing onsite contact interaction via $s$-wave scattering. Interactions can also lead to depopulation of the Bloch band \cite{muller2007state}. A strong interaction couples to excited bands. This leads to renormalization of parameters of the tight-binding Hamiltonian \cite{johnson2009effective,mering2011multiband,luhmann2012multi}. Also, several early works on optical lattices investigated the interaction stability of the system bands, including excited bands \cite{isacsson2005multiflavor,muller2007state,wirth2011evidence}. For example, quasi-resonant two-body processes allow for depopulation of the $p$-band via the process $p+p\to s+d$. For a review, see \cite{li2016physics}.

The AC-Stark shift-based implementation of the optical potential achieves a spatial resolution of at most half of the wavelength of the laser used. The work \cite{Lacki2016} proposes a different method based on the dark state in the three-level 
$\Lambda$-system (for a broader context, see \cite{dalibard2011colloquium}). Then, the atomic potential consists of subwavelength-width peaks. An experimental implementation of the scheme \cite{Yang2018} completely avoided contact interactions using a fermionic isotope of ytterbium. However, when the dark-state potential is applied to interacting bosons, the stability of, for example, its lowest Bloch band against the contact interactions must be carefully considered.

In this work, we use the Hamiltonian
\begin{align}
H_D=\!\!\int\!\!\psi^\dagger(\vec r)\!\left[-\frac{\hbar^2}{2m}\frac{d^2}{dx^2}\!+\!V_{na,D}(x)\!\right]\!\psi_\sigma(\vec r) d^3r\!+\!H_{\textrm{int.},D}, 
\label{eqn:hamH0}
\end{align}
for interacting bosons in the lowest Bloch band of the dark state $V_{na,D}(x)$ potential, interacting via $H_{\textrm{int.},D}$. Section \ref{section:model} discusses the one-particle foundational model. There, Section \ref{subsec:decompHa} describes the decomposition of the $\Lambda$ system into dark state channel with zero energy and two bright-state channels, one with a large positive energy, and the other with a large negative energy. 
The properties of the three channels are discussed in \ref{sec:spchannels}. 
The interacting system is analyzed in Section \ref{subsec:2pint}. In \ref{subsec:2pBH}, we review a perfect 2-particle dark state with no loses, as described in \cite{praca2008}. Interaction depletes the dark state channel by promoting one of the particles to the negative energy bright state channel, and the other to a high energy state. This is treated using Fermi's Golden Rule (FGR) in subsequent sections of \ref{subsec:2pint} beginning with~\ref{subsec:fgrtheory}. The main result is estimation of losses  for a weakly interacting dark state gas within the $\Lambda$ system. This is obtained numerically in \ref{subsec:fgrnumeric}, \ref{subsec:transmodes}, \ref{subsection:2pBHint2}, and \ref{subsec:avglossrate}. The conclusions are made in \ref{sec:conclusions}.

\section{Model}
\label{section:model}
We consider a bosonic gas populating a three level system described by the following Hamiltonian:
\begin{eqnarray}
H&=&\!\!\!\!\!\!\!\!\sum_{\sigma=g_1,e,g_2}\!\!\!\!\!\!\!\!\int\!\! \psi_\sigma^\dagger(\vec r) [H_1(\vec r)\!\!+\!\!V_{\textrm{tr}}(\vec r)]\psi_\sigma(\vec r) d^3r \!+\! H_{\textrm{int}.}, \\
H_1(\vec r)&=&T+H_a(x),\quad T=-\frac{\hbar^2}{2m}\nabla^2,\\
H_{1,x}(x)&=&T_x+H_a(x),\quad T_x=-\frac{\hbar^2}{2m}\frac{\partial^2}{\partial x^2},\\
\vec{r}&=&(x,y,z),\nabla=(\partial_x,\partial_y,\partial_z).\nonumber
\label{eqn:hamiltonianH}
\end{eqnarray}
The $V_{\textrm{tr}}(\vec r)=\frac{1}{2}m\omega_\perp^2(y^2+z^2)$ confines the atoms along the $x$ axis. The Hamiltonian in the $y,z$ directions is a harmonic oscillator Hamiltonian with eigenfunctions $H_\alpha(y) H_\beta(z),\alpha\in \mathbb{N}$. If the gas is restricted to the lowest transverse mode, one may consider a 1D restriction of \eqref{eqn:hamiltonianH}. This is especially suited for the non-interacting gas, as the only term of \eqref{eqn:hamiltonianH} that can couple different transverse modes is the interaction term $H_{\textrm{int}.}$. For this purpose, we define $x$-restricted $H_{1,x}$ and $T_x$.

\begin{figure}
    \centering
    \includegraphics[width=8.4cm]{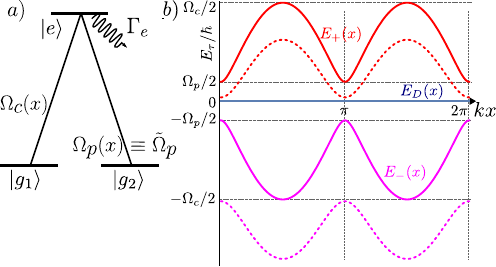}
    \caption{Panel a) the three level $\Lambda$ scheme used in this work, the $g_1, g_2$ denote the stable hyperfine states, and $e$ is an excited state coupled optically via lasers with Rabi frequencies $\Omega_c(x)$ (position dependent, sine standing wave -- see text) and a constant $\Omega_p=\tilde{\Omega}_p$.  Panel b) the eigenenergies $E_\tau(x)$ of the Hamiltonian $H_a(x)$ [Eq.~\eqref{eqn:3levels} and \eqref{eq:3evelEplus}] for $\tilde{\Omega}_p=0.2\tilde{\Omega}_c, \Delta=0,\Gamma_e=0$ (solid lines) and $\tilde{\Omega}_p=0.2\tilde{\Omega}_c, \Delta=0.5\tilde{\Omega}_c,\Gamma_e=0$ (dashed lines). }
    \label{fig:intro}
\end{figure}

\paragraph*{The term $H_a$.}
The $H_a$ describes coupling of two stable hyperfine states $|g_1\rangle, |g_2\rangle$ and $|e\rangle $ -- the excited atomic state with a spontaneous emission rate $\Gamma_e$, and detuning $\Delta$, coupled as in  Fig.~\ref{fig:intro}(a):
\begin{equation}
H_a(x)= \frac{\hbar}{2} \left(\begin{array}{ccc} 0 & \Omega_c(x) & 0 \\ \Omega_c(x) & -i\Gamma_e -2\Delta & \Omega_p(x) \\ 0 & \Omega_p(x) & 0  \end{array}\right).
\label{eqn:3levels}
\end{equation}
The Rabi frequencies $\Omega_c(x)$ and $\Omega_p(x)$ we consider are as in~\cite{Lacki2016}:
\begin{eqnarray}
    \Omega_c(x)&=&\tilde{\Omega}_c \sin(kx),\\
    \Omega_p(x)&=&\tilde{\Omega}_p \equiv \textrm{const.}
    \label{eqn:Omegas}
\end{eqnarray}
The $k$ is the wave number of the laser standing wave creating $\Omega_c(x)$,  $k=2\pi /\lambda_L$. 
The $k$ defines a recoil energy unit: 
\begin{equation}
    E_R=\frac{\hbar^2 k^2}{2m}.
\end{equation}
We will consider $\tilde{\Omega}_c,\tilde{\Omega}_p $ in thousands of $ E_R$. By incorporating all phase factors into $|g_i\rangle,|e\rangle$, we can assume that $\tilde{\Omega}_c,\tilde{\Omega}_p\in\mathbb{R}$.

\paragraph*{The interaction Hamiltonian $H_\textrm{int.}$.} The Hamiltonian $H_\textrm{int.}$ describes contact interactions between atoms. We assume it to be of the form 
\begin{equation}
    H_{\textrm{int}}\! =\!\! \frac{1}{2}\!\!\int\!\!\dd{\vec{r}_1}\dd{\vec{r}_2}
    \hat{\psi}_\sigma^\dag(\vec{r}_1)
    \hat{\psi}_\tau^\dag(\vec{r}_2)
    g_{3D}^{\sigma\tau\rho\eta}(\vec{r}_1, \vec{r}_2)
    \hat{\psi}_\rho(\vec{r}_2)
    \hat{\psi}_\eta(\vec{r}_1),
\label{eq:V3D}       
\end{equation}
with the scattering tensor
\begin{equation}
    g_{3D}^{\sigma\tau\rho\eta}(\vec{r}_1, \vec{r}_2) = g_{3D,0} \delta_{\sigma \eta}\delta_{\tau \rho}\delta(\vec{r}_1-\vec{r}_2),
    \label{eq:gtensordelta}
\end{equation}
which forbids changes of the internal state during collision and is characterized by the same scattering length in each channel. Consequently, the interaction term simplifies to
\begin{equation}
     H_{\textrm{int}} = \frac{g_{3D,0}}{2} \int\dd{\vec{r}} \hat{\psi}_\sigma^\dag(\vec{r})\hat{\psi}_\tau^\dag(\vec{r})\hat{\psi}_\tau(\vec{r})\hat{\psi}_\sigma(\vec{r}).
     \label{eq:2pinthint}
\end{equation}
We also define the 1D projection, $g_{1D}$ of the interaction Hamiltonian, whose coeffients satisfy
\begin{equation}
    g_{1D}^{\sigma\tau\rho\eta}(\vec{r}_1, \vec{r}_2)\delta(y_1-y_2)\delta(z_1-z_2)=g_{3D}^{\sigma\tau\rho\eta}(\vec{r}_1, \vec{r}_2).
    \label{eq:gtensordelta1d}
\end{equation}
and the 1D projection of the $H_{\textrm{int}}:$
\begin{equation}
    \hat{V}_1\! =\!\! \frac{1}{2}\!\!\int\!\!\dd{x_1}\dd{x_2}
    \hat{\psi}_\sigma^\dag(x_1)
    \hat{\psi}_\tau^\dag(x_2)
    g_{1D}^{\sigma\tau\rho\eta}(x_1, x_2)
    \hat{\psi}_\rho(x_2)
    \hat{\psi}_\eta(x_1),
\label{eq:V1D}    
\end{equation}
The interactions are further discussed in Section \ref{subsec:2pint}. 

We first discuss the decomposition of the single particle $H_a$ Hamiltonian into two dark  and two bright state channels.

\subsection{Decomposition of \texorpdfstring{$H_a$}{H\_a}}
\label{subsec:decompHa}
We summarize the decomposition of $H_a$ as well as $H_{1,x}$ and $H_{1}$ into dark and bright state channels.  It has been described in detail in~\cite{Lacki2016}. 

The Hamiltonian $H_a(x)$ for each $x$ has a ``dark state'' eigenvector:
\begin{eqnarray}
|D(x)\rangle & = & \frac{-\Omega_{c}(x)|g_2\rangle+\Omega_{p}|g_1\rangle}{\sqrt{|\Omega_{p}|^{2}+|\Omega_{c}(x)|^{2}}},\label{eq:darkState3}
\end{eqnarray}
to a zero eigenvalue $E_D(x)=0$. The two other bright state eigenvectors $|B_\pm(x)\rangle$, with a nonzero contribution of an excited state $|e\rangle $, have the energies:
\begin{equation}
E_\pm(x)=\frac{\hbar}{4}\left(-i\tilde{\Gamma}\pm\sqrt{-\tilde{\Gamma}_{e}^{2}+4|\tilde{\Omega}_{p}|^{2}+4|\Omega_{c}(x)|^{2}}\right),\label{eq:3evelEplus}
\end{equation}
with $\tilde{\Gamma}=\Gamma_e+2i\Delta$. The dependence of eigenenergies on the spatial component $x$ is shown in Fig.~\ref{fig:intro}(b).

A set 
\begin{equation}
 {\cal B}=\{|D(x)\rangle,|B_{+}(x)\rangle,|B_{-}(x)\rangle\}= \{| { \cal{B} }_\tau \rangle, \tau  \in \{D,B\pm \} \} ,
\end{equation}
complemented with a biorthonormal left eigenvectors $\langle D(x)|,\langle B_{+}(x)|,\langle B_{-}(x)|$ can be used to decompose the $H_{1,x}$ Hamiltonian by projecting:
\begin{equation}
    H_{1,\tau}=P_\tau H_{1,x} P_\tau,\quad\tau\in\{D,B_+,B_-\}
    \label{eq:hamiltonianSingleChannelProjection}
\end{equation}
where $P_\tau=\sum_{i=1}^3 |{\cal B}_\tau\rangle \langle {\cal B}_\tau |,\ \tau\in\{D,B_+,B_-\}$.
The $H_{1,\tau}$ describes a particle moving in a scalar potential $V_{na,\tau}$:
\begin{equation}
    H_{1,\tau}\!=\!\!\int\!\!\psi_{\tau}^\dagger(x)\!\left(\! -\frac{\hbar^2}{2m}\frac{d^2}{dx^2}\!+\!V_{na,\tau}(x)\!+\!E_\tau(x)\!\!\right)\!\psi_{\tau}(x) dx,
    \label{eqn:hamiltonianSingleChannelGeneric}
\end{equation}
where 
\begin{equation}
    V_{na,\tau}(x)=-\Bigl\langle {\cal B}_\tau (x)\Bigl| \frac{d^2}{dx^2} \Bigl| {\cal B}_\tau (x)\Bigr\rangle.
\end{equation}
Although the fundamental period of the above is $a=\lambda_L/2$, we treat it as a $\lambda_L$-periodic potential. This corresponds to the period of $H_a$.

The corresponding Bloch eigenfunctions and quasienergies $E_{n,q,\tau}$ are indexed by $n$:
\begin{equation}
  \psi_{n,q,\tau}(x) = {\cal N} u_{n,q,\tau}(x) e^{iqx/\hbar }.
  \label{eqn:Blochsingle}
\end{equation}
where $u_{n,q,\tau}(x)$ is an $\lambda_L/2$-periodic function, and ${\cal N}$ is a normalizing factor.
We assume a natural ordering in which the eigenvalues increase with the index $n$. 
The quasimomentum $q$ lies within the Brillouin zone,  BZ=$[-\pi/\lambda_L,\pi/\lambda_L)$ \eqref{eqn:3levels}.  We demand that the periodic part of the Bloch function satisfies the condition $\int_0^{a
}|u_{n,q,\tau}(x)|^2 \dd{x}=1$.

\subsection{Single particle channels}
\label{sec:spchannels}
\begin{figure}
    \centering
    \includegraphics[width=8.4cm]{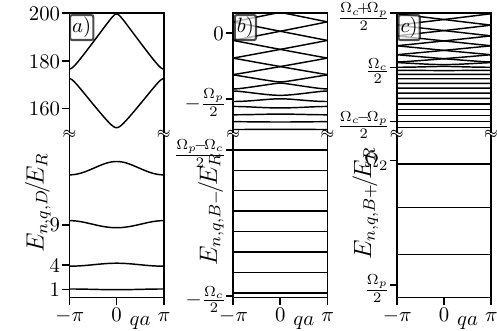}
    \caption{The Bloch quasi-energies of dark and bright state channels. Panel a) shows the dark state channel with potential $V_{na,D}(x)$ -- Eq.~\eqref{eqn:VnaD}. Panels b) and c) show bright state channels $B_-$ and $B_+$ -- Eq.~\eqref{eq:3evelEplus}. } 
    \label{fig:singleparticlechannels}
\end{figure}

\paragraph*{Dark state channels.}The dark state potential, for $\Omega_p(x), \Omega_c(x)$  as in~\eqref{eqn:Omegas} has the form:
\begin{equation}
V_{na,D}(x)=\frac{\epsilon^2 \cos^2 kx}{\qty(\epsilon^2+\sin^2 kx)^2},\ \ \epsilon=\frac{\tilde{\Omega}_p}{\tilde{\Omega}_c}.
\label{eqn:VnaD}
\end{equation}
It consists of deep potential wells separated by periodically arranged narrow peaks. The low-energy Bloch spectrum resembles that of a ``particle in the box'' $\bar{E}_n \approx n^2 E_R$ for small $n$. For energies greater than $E_R/\epsilon^2$ the band gaps decrease exponentially with band number, and the energy level structure transitions to that of a free particle . This behavior is illustrated in Fig.\ref{fig:singleparticlechannels}a) where we show the Bloch spectrum for $\epsilon=0.1$ and energies $E=0 \olddiv 200 E_R =2 E_R / \epsilon^2$.

\paragraph*{Bright state channels.} For bright state channels the scalar potential in the Hamiltonian \eqref{eqn:hamiltonianSingleChannelGeneric} 
is $E_{\pm}(x) + V_{na,B_\pm}$ with $E_\pm(x)$ being a dominant term [see Fig. \ref{fig:singleparticlechannels}b,c)]. For $\Delta=0$, the potential minima/maxima $E_-(x)$ are at $- \tilde{\Omega}_c/2$ or $-\tilde{\Omega}_p/2$, respectively. The $E_+(x)$ has a minimum at $\tilde{\Omega}_p/2$ and a maximum at $\tilde{\Omega}_c/2$. In both cases for energies below the potential maximum the bands are nearly equidistant and flat akin to harmonic oscillator levels. The spectrum above the maximum of $E_{\pm}(x)$ resembles that of a free particle. 

We also observe that the bright states spectrum $B_-$ near zero is already highly excited and $\approx \tilde{\Omega}_p$ (for $\Delta=0$) above the threshold $V_{na,B_\pm}$. The near-parabolic dependence of $E_{n,q,B-}$ high above the threshold of $E_-(x)$ implies that $|\frac{d}{dq}E_{n,q,B-}|$ is very large.

\paragraph*{Single particle losses.} The two bright and dark state channels, described by the Hamiltonian $H_{1,\tau}$,~Eq.\eqref{eq:hamiltonianSingleChannelProjection} are coupled by the terms contained in $H_1(\vec r)$ that are not included in each $H_{1,\tau}$. The effect of those terms is primarily introduction of a small imaginary part of $E_{n,q,
\tau}$ which is interpreted as the finite lifetime of Bloch states due to $[T,H_a(x)]\neq 0$. This has been systematically studied in see~\cite{Lacki2016}. In this work, we focus on contribution to the finite lifetime due to interactions.

\section{Two-particle interacting system}
\label{subsec:2pint}
Here we consider a case of two interacting atoms. In Section~\ref{subsec:2pBH} we describe the dynamics of two atoms within the dark state channel, the lowest Bloch band of the $V_{na,D}(x)$ potential, and the lowest transverse harmonic modes of $V_{tr}(\vec r)$. This defines a subspace of the entire Hilbert space for Eq.\eqref{eqn:hamiltonianH}. The stability of this subspace and the associated loss rates are studied in the subsequent sections \ref{subsec:fgrtheory}, \ref{subsec:fgrnumeric}, and \ref{subsec:transmodes} using the FGR. The final numerical analysis of losses is presented in Sections \ref{subsection:2pBHint2} and \ref{subsec:avglossrate}.

\subsection{Two particle Bose-Hubbard model}
\label{subsec:2pBH}

For two atoms populating the lowest transverse bands and the lowest band in the $V_{na,D}$ potential, the interactions originally present in $H$ as $H_\textrm{int.}$ may be treated by a tight-binding approach. To this end, we use the Wannier function expansion of the field operator $\psi(\vec r)$  -- see \cite{Kohn1959,Jaksch2005}. This leads to a Bose-Hubbard model \cite{Jaksch1998}:
\begin{equation}
 H_{BH}=-J\sum_i a_i^\dagger a_{i+1} + a_{i+1}^\dagger a_{i} + \frac{U}{2}\sum_i n_i(n_i-1),
\label{eqn:bh}
\end{equation}
with 
\begin{equation}
J=-\int W(x) \left(-\frac{\hbar^2}{2m}\frac{d^2}{dx^2}+V_{na,D}(x)\right)W\left(x+a\right) \textrm{d}x,
\end{equation}
and 
\begin{eqnarray}
U=&& \underbrace{\frac{g_{3D,0}}{2}\int H^4_0(y)H^4_0(z) \dd{y}\dd{z}}_{{g}_0}\int\!\!  |W(x)|^4 \dd{x}. \\
\label{eqn:Uform}
\end{eqnarray}
The value of $J$ can be shown for $\epsilon\ll 1$ to be $J=\frac{2E_R\epsilon}{\pi^2}$\cite{Lacki2016}. The $U$, in the simplest case, for $g_{3D}^{\sigma\tau\rho\eta}$ given by Eq.~\eqref{eq:gtensordelta}, for $\epsilon\ll 1$ is approximately $U\approx \frac{1}{2} {\frac{m\omega}{2\hbar\pi}} \frac{3g_0}{2a}$.
The geometry of the transverse trap, in particular the value of $\omega_\perp$, affects the value of $g_{0}$, an effective interaction strength.

The eigenvectors of the single-particle part of \eqref{eqn:bh} are plane waves indexed by $q\in [-2\pi/\lambda_L,2\pi/\lambda_L)$: 

\begin{align}
    E_q &= -2J \cos(qa), \\
    \psi_q(x) &= N_q \sum_{j} e^{ijqa} W(x-ja).
    \label{eqn:dimerJ}
\end{align}
where $N_q$ is a normalization factor.
The Hamiltonian \eqref{eqn:bh} for a lattice with a total of two particles was investigated in~\cite{Valiente2008}. The authors show that there are two distinct classes of eigenvectors, whose eigenvalues are separated by an energy gap $\sim U$, for large $U$.

The first one encompasses quasi-free states $\phi^{qf}_{q_1,q_2}$ with a spectrum resembling the non-interacting case:
\begin{equation} \label{eq:E_free}
    E_{q_1,q_2} = E_{Q,q} = -2J_Q \cos(q a).
\end{equation}
The amplitude $J_Q = 2J \cos(Qa/2)$ depends on the total quasimomentum $Q$, preserved by the contact interactions, and the parameter $q$. In case of $U=0$ we have that  
\begin{equation} 
q = (q_1 - q_2)/2.
\label{eqn:qhalg}
\end{equation}
For finite interaction, $U\neq 0$, the equality \eqref{eq:E_free} holds, but the quantum number $q$ is no longer given by \eqref{eqn:qhalg}. The values of $q$ for $U\neq 0$ continue to densely fill $[-\pi/a,\pi/a)$. As a consequence, for a fixed $Q$, the $E_{Q,q}$ take all possible values between $-2 J_Q$ and $2J_Q$. In this work, we do not directly calculate $q$, but we refer to different states with the same $Q$ by their energy.

The second class are interaction-bound dimer-like states $\phi^{\textrm{dimer}}_{Q}$, which are localized in the relative position coordinates around $x_1 - x_2=0$ , with the energies
\begin{equation}
    E_{Q} = \sign(U)\sqrt{U^2 + 4J_Q^2}.
    \label{eq:twoparticlestrip}
\end{equation}
For $U > 0$, the dimer energies are above the quasi-free spectrum \eqref{eq:E_free}, while for $U < 0$ they are below. 

It is worth noting that the dimer-like states exist for both attractive and repulsive interactions. In a periodic lattice with $N$ sites, there are $N(N-1)/2 \sim N^2$ quasi-free and $N$ dimer-like states.

\subsection{Losses from two-particle dark state: Fermi's Golden Rule}
\label{subsec:fgrtheory}
We are now going to introduce a notation useful to navigate among the various states of $\Lambda$-system including the transverse modes -- different states that can be populated by means of the contact interactions.

The $y,z$ directions in $H_{1}(\vec r)+V_{tr}(\vec r)$ Hamiltonian can be factored out and the eigenstates of that two particle states of that Hamiltonian are of the form 
\begin{equation}
    |\phi;  \bm{\alpha},\bm{\beta}\rangle = |\phi\rangle |(\alpha_1,\beta_1),(\alpha_2,\beta_2)\rangle,
    \label{eqs:state}
\end{equation}
where the indices $\bm\alpha=(\alpha_1,\alpha_2),\bm\beta=(\beta_1,\beta_2)$ indicate populated transverse modes. The $|\phi\rangle$ are the two-particle states of $H_{1,x}$. We are particularly interested in states with $\bm{\alpha},\bm{\beta}=0$ where $|\psi\rangle$ is dark-state only and $|\phi\rangle \equiv \phi(x_1,x_2)=\psi^{\textrm{dimer}}(x_1,x_2)$ or $\psi^{qf}(x_1,x_2)$. 
This is a set of the ``initial'' states whose collisional stability is analyzed with the FGR.

The ``final'' states are also of the form \eqref{eqs:state}, but $|\phi\rangle$ is a two-particle product of single-particle eigenstates of $H_{1,D}, H_{1,B+}$ or $H_{1,B-}$. At least one of the atoms occupies the $B_+$ or $B_-$ manifold. Also. in general $\bm\alpha,\bm\beta\neq0$. All single-particle states have the form of Eq.~\eqref{eqn:Blochsingle}. These states are indexed by quasimomenta $q_1,q_2$. The contact interaction preserves the total quasimomentum, leaving a single continuum parameter, just as in Eq.~\eqref{eqn:qhalg}, which parameterizes the continuum of states to which a particular two-particle ``initial state'' is coupled. In addition, the two-particle states are indexed by the eigenvalue numbers $n$ and $\tau\in\{D,B+,B-\}$. These discrete parameters, together with $\bm{\alpha},\bm{\beta}$, are then summed to obtain the total decay rate. 

\paragraph*{Fermi's Golden Rule} 
The transition probability from a discrete state $\ket{\phi_0}$ with energy $E$ into a continuum of states $|\chi_u\rangle$ with energies $E_u$ indexed by $u$ is given by the FGR:
\begin{equation}\label{eq:fgrfinal}
    \Gamma = \frac{2\pi}{\hbar} \overline{\abs{\mel{\chi_{u_0}}{\hat{V}_0}{\phi_0}}^2 \rho(E)}.
\end{equation}
Here, the state $\ket{\phi_0}$ is a two-particle ``initial'' state of the form \eqref{eqs:state}, with $\bm\alpha = \bm\beta = 0$, and a weak interaction potential $\hat{V}_0$ and $E_{u_ 0}=E$ was assumed. The overline $\overline{\ \square\ }$ denotes the summation over all decay channels. Moreover, $\rho(E)\delta E$ is a total number of states with energy in the interval $(E,E+\delta E)$ (not per unit volume). 

In application of the FGR, we normalize the final states to 1. To this end, we work in a finite-size box of length $L=Na$ under periodic boundary conditions.

\subsection{The Fermi's Golden Rule -- numerics}
\label{subsec:fgrnumeric}
We first consider the decay of an initial state $|\phi;\bm{\alpha}=0,\bm{\beta}=0\rangle$ with an energy close $2E_R$ and a total quasimomentum $Q=q_1+q_2$. This state includes interactions as described in \ref{subsec:2pBH}. 

We now consider decay into an individual decay channel. To do this, we fix the final transverse mode configuration $\bm{\alpha}_f, \bm{\beta}_f$ and the eigenvalue index of the $H_{1,x}$ Hamiltonian as described in the preceding Section. The target state is also of form \eqref{eqs:state}, but its $x$-part is a product state $|\psi_1,\psi_2\rangle $ with $\psi_i \equiv \psi_{n_i,q_i,\tau_i}$ [Eq.~\eqref{eqn:Blochsingle}] as we do not pre-diagonalize the interactions in the final state subspace.

By FGR, the single channel (sc) decay rate is:
\begin{align}
    \Gamma_{\bm\alpha_f,\bm\beta_f}^{sc} = 
        &\frac{2\pi}{\hbar}
        \int\limits_{[-\frac{\pi}{a},\frac{\pi}{a}]^2} \dd{q_1}\dd{q_2} |M_{\bm\alpha_f,\bm\beta_f}|^2
        \abs{\mel{\psi_1,\psi_2}{\hat{V}_1}{\phi}}^2 \times \notag \\ 
         & \!\!\!\!\!\!\!\!\!\!\!\!\!  \times\frac{L}{2}\delta(Q\!-\!q_1\! - q_2)\rho^{\bm{q}}\left[E_0\!-\!\hbar\omega_\perp(\norm{\bm{\alpha}_f\!}+\norm{\bm{\beta}_f}\!)\right], \label{eqn:gammaab} \\
\rho^{\bm{q}}(E) & =\delta\left[E-\sum_{i=1}^2 E_{n_i,q_i,\tau_i}\!\right],\\
    \psi_i &\equiv \psi_{n_i,q_i,\tau_i}, \\
    M_{\bm{\alpha}_f,\bm{\beta}_f} =& \int  H_{\alpha_1^f}(y)H_{\beta_1^f}(y)H_{\alpha_2^f}(z)H_{\beta_2^f}(z) \times \notag \\
    & \ \ \ \ \ \ \ \ \ \ \ \ \ \ \ \times H_0^2(y)H_0^2(z) \dd{y}\dd{z} .
\end{align}
where $\norm{\bm{\alpha}_f}=\alpha^f_1+\alpha^f_2, \norm{\bm{\beta}_f}=\beta^f_1+\beta^f_2 $.
The $\Gamma_{\bm{\alpha}_f,\bm{\beta}}^{sc}$ depends on $Q,n_1,n_2,\tau_1,\tau_2$. We omit them from the symbol for the simplicity of notation. 
The $\delta(Q - q_1 - q_2)$ is due to the quasimomentum conservation.  The density of states (DOS) of a final product state can be expressed as

\begin{align}
    \rho^{\bm{q}}(E_0)
    =&
        \frac{L}{2}\sum_{\substack{q_1, q_2: \\ q_1 + q_2 = Q,\\\norm{E}=E_0}} \frac{1}{\abs{\dv{E_{n_1,q_1,\tau_1}}{q_1} - \dv{E_{n_2,q_2,\tau_2}}{q_2}}},\\
        \norm{E}=&E_{n_1,q_1,\tau_1}+E_{n_2,q_2,\tau_2}. \notag \\
        \label{eqn:densityOfStates}
\end{align}
We can write the FGR expression for a single-channel loss rate as
\begin{eqnarray}
    \Gamma_{{\bm{\alpha}}_f,{\bm{\beta}}_f}^{sc} &=& \frac{L\pi}{\hbar}
    {|M_{\bm\alpha_f\bm\beta_f}|^2\abs{\mel{\psi_1,\psi_2}{\hat{V}_1}{\phi}}^2  \times}\notag \\
        && {  \times   \rho^{\bm{q}}\left[(  E_0-\hbar\omega_\perp( \norm{\bm{\alpha}} + \norm{\bm{\beta}} )\right]. }
\label{eqn:gammaab2}
\end{eqnarray}
The two-particle state is a normalized product of Bloch states $\psi_i\equiv \psi_{n_i,q_i,\tau_i}$:
\begin{align}
    \langle x\ket{\psi_1,\psi_2}\equiv&
        \frac{1}{\sqrt{S_{\psi_1,\psi_2}}} \sum_{i, j}\qty(\psi_1^i(x) \psi_2^j(x) + \psi_2^i(x) \psi_1^j(x)) .
\end{align}
The factor $S_{\psi_1,\psi_2},$  comes from quantum statistics and is equal to 1 if both $\ket{\psi_1},\ket{\psi_2}$ are different and 2 if they are the same.

By expanding the initial state $\ket{\phi}$:
\begin{equation}
    \ket{\phi} = \sum_{q_3 + q_4 = Q} v_{q_3 q_4} \ket{\psi_{0,q_3,D}, \psi_{0,q_4,D}},
    \label{eqn:vqq_coeffs}
\end{equation}
where $\psi_3\equiv \psi_{0,q_3,D}, \psi_2 \equiv \psi_{0,q_4,D}$, we have that
\begin{align}
    \mel{\psi_1,\psi_2}{\hat{V}_1}{\phi} &= \frac{g_{3D,0}}{2L} \sum_{q_3 + q_4 = Q} v_{q_3 q_4} V_{q_1,q_2,q_3,q_4}^{n_1 n_2} \notag \\
    &\equiv \frac{g_{3D,0}}{2L} \mathcal{F}^{n_1,n_2}_{q_1,q_2}, \label{eqn:matel} \\
      V_{q_1 q_2 q_3 q_4}^{n_1 n_2}  =&
        \frac{a \delta_{q_1+q_2,q_3+q_4} }{\sqrt{S_{\psi_1,\psi_2} S_{\psi_{0,q_3,D},\psi_{0,q_4,D}}}} \sum_{i, j} \int_{0}^{a}\dd{x} \times \notag \\
        & \times \qty(u_1^i(x) u_2^j(x) + u_2^i(x) u_1^j(x))^\dag \times \notag \\
        & \times \qty(u_3^j(x) u_4^i(x) + u_4^j(x) u_3^i(x)).  \notag \\
\end{align}
where $u^i_n$ is the $i$-th component ($g_1, g_2 $ or $e$) of the periodic part of $\psi_n$.

In addition to conserving the quasimomentum $q_1 + q_2 = Q = q_3 + q_4 $, the bands $n_1$, $n_2$ of the final states must be chosen in a way that the total energy $E$ is also conserved.

\subsection{Summation and averaging of loss rates transverse modes}
\label{subsec:transmodes}
The total decay rate is given by summing over all decay channels: defined by the index $\tau,n $ in Eq.~\ref{eqn:Blochsingle} and over all transverse modes $\bm{\alpha}_f,\bm{\beta}_f$ as in Eq.~\eqref{eqn:gammaab}. 
In what follows, we will consider partial sums or averages of $\Gamma_{{\bm{\alpha}_f,\bm{\beta}_f}}^{sc}$ over a subset of parameters of $Q,n_1,n_2,\tau_1,\tau_2$. When the loss rate is summed over all possible final states indexes $n_1,n_2$ we obtain
\begin{equation}
\Gamma_{\bm{\alpha}_f,\bm{\beta}_f}^{Q,\tau_1,\tau_2} = \sum\limits_{n_1,n_2} \Gamma_{\bm{\alpha}_f,\bm{\beta}_f}^{sc}
\label{eq:sumsc}
\end{equation}
We call $\Gamma_{\textrm{tot}}$ the sum of the above over $\bm{\alpha}_f,\bm{\beta}_f$:
\begin{equation}
\Gamma^{Q,\tau_1,\tau_2}_\textrm{tot}  = \sum\limits_{{\bm\alpha}_f,{\bm\beta}_f} \Gamma_{\bm{\alpha}_f,\bm{\beta}_f}^{Q,\tau_1,\tau_2}
\end{equation}
The latter sum can be simplified by nothing that:
\begin{equation}
\sum\limits_{\bm{\alpha}_f,\bm{\beta}_f} \Gamma_{\bm{\alpha}_f,\bm{\beta}_f}^{Q,\tau_1,\tau_2}= \sum\limits_{n=0}^\infty \sum\limits_{\substack{\bm{\alpha}_f,\bm{\beta}_f: \\ \norm{\bm{\alpha}_f}\!+\!\norm{\bm{\beta}_f}=n }} \Gamma_{\bm{\alpha}_f,\bm{\beta}_f}^{Q,\tau_1,\tau_2}.
\end{equation}
Expanding the right side, by going back to $\Gamma_{\bm{\alpha}_f,\bm{\beta}_f}^{sc}$ using Eqs.~\eqref{eq:sumsc}-~\ref{eqn:gammaab2}, we have:
\begin{align}
\sum\limits_{\bm{\alpha}_f,\bm{\beta}_f}& \Gamma_{\bm{\alpha}_f,\bm{\beta}_f}^{Q,\tau_1,\tau_2}=\frac{L\pi}{\hbar}\abs{\mel{\psi_1,\psi_2}{\hat{V}_1}{\phi}}^2\times \notag \\
& \times \sum\limits_{n=0}^\infty  \rho^{\bm{q}}( E_0\!-\!n\hbar\omega_\perp\!)
\underbrace{\sum\limits_{\substack{\bm{\alpha}_f,\bm{\beta}_f: \\ \norm{\bm{\alpha}_f}\!+\!\norm{\bm{\beta}_f}=n }} \!\!\!\!\!\!\!\!\!\!\!|M_{\bm\alpha_f\bm\beta_f}|^2}_{S_n}.
\label{eqn:transverseSummation}
\end{align}
One can show that $S_{2k+1}=0, k\in \mathbb{N}$ and that $S_{2k}=S_0,k\in\mathbb{N}$.

The averages of $\Gamma$ are taken with respect to a different initial states, for example: 
\begin{equation}
\left\langle \Gamma^{sc}_{{\bm\alpha}_f,{\bm\beta}_f} \right\rangle = \frac{1}{N^2} \sum\limits_{\phi\in\textrm{dimer}\cup{\textrm{qf}}} \Gamma_{\bm{\alpha}_f,\bm{\beta}_f}^{sc},
\end{equation} where ``dimer'' and ``qf'' refer to a set of $N$ dimer and $N(N-1)$ quasi-free states. We will also use $\left\langle \Gamma^{\tau_1,\tau_2}_\textrm{tot} \right\rangle $ which averages $\Gamma^{Q,\tau_1,\tau_2}_\textrm{tot}$ over different initial states $\phi$. When the mean is taken with respect to just the dimer or the quasi-free states we use $\left\langle \,\cdot\, \right\rangle_{\textrm{dimer}}$ and $\left\langle \,\cdot\, \right\rangle_{\textrm{qf}}$, respectively.

We can also write a closed formula for the total $\langle\Gamma_\textrm{tot}\rangle$ where the overbar $\overline{\ \cdot\ }$ is the summation and averaging over all the above parameters, and finally over all channels $\tau_1,\tau_2\in \{D,B_+,B_-\}$:
\begin{equation}
    \langle\Gamma_\textrm{tot}\rangle = \frac{\pi g_{3D,0}^2}{\hbar L^2} \overline{\abs{\mathcal{F}}^2 \rho^{\bm{q}}(E_0)}.
    \label{eq:gammafinal}
\end{equation}

The loss rate depends on the size of the system. For a system of two particles in the $L$-site lattice, the particle density decreases in the $L\to\infty$ limit. This reduces the chances for the two particles to interact, and, consequently, it lowers the loss rate. 

A meaningful limit can be obtained if the system is assumed to be prepared in a product state (for weak interactions) of $N=\nu L$ particles, where $\nu$ is the filling factor. In \cite{li2016physics} a similar situation of depletion of a metastable lattice band is discussed (our band is metastable due to the existence of bright state channels, despite being the $s$ band within the dark state channel). 

By analogy, if we assume that the state of the system is approximated by an initial condensate state of the form $\frac{1}{N!} a_{n=0,q=0,D}^\dagger |\textrm{vac}\rangle$, then the decay resulting from the interaction-driven depletion of the dark state band occurs by the interaction of any pair of atoms out of a total of $N$. This gives an additional $N(N-1)$ factor to the total loss rate. The loss rate per site is thus $\sim \nu^2 N \langle\tilde{\Gamma}_\textrm{tot}\rangle$. This quantity does not depend on the size of the system, and this is the quantity that we plot in the following Section for $\nu=1$. 
\begin{figure}[b]
    \centering
    \includegraphics[width=0.95\linewidth]{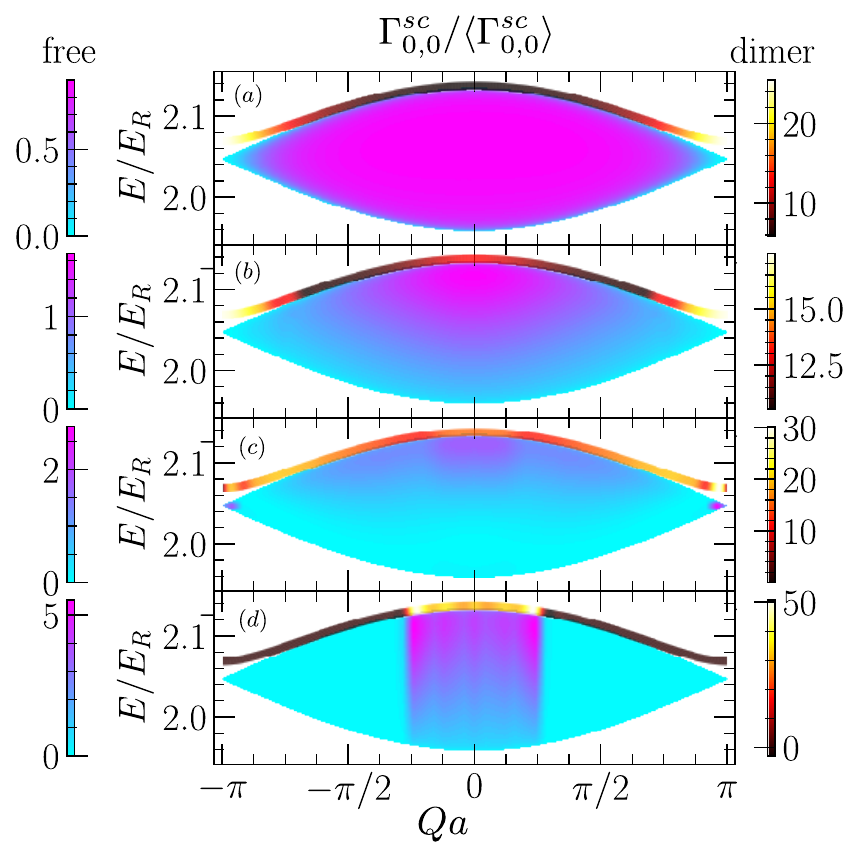}
    \caption{The dependence of $\Gamma^{sc}_{\bm\alpha_f=0,\bm\beta_f=0}$ on the total quasimomentum $Q=q_1+q_2$ and energy $E$ of the dark initial state $|\phi_0\rangle $, for $N = 50$ lattice sites, $U/t = 1$ ($g_0 \approx 0.0146$), $\tilde{\Omega}_c = 4 \times 10^4 E_R=10\tilde{\Omega}_p$ for the decay channel $\tau_1=D,\tau_2=B_-$, where $n_1$ indicates the eigenvalue number in the final dark state.  (a) $n_1 = 0, 
    \langle{\Gamma^{sc}_{0,0}}\rangle \approx 3.14 \times 10^{-11} E_R$, (b) $n_1 = 6, 
    \langle{\Gamma^{sc}_{0,0}}\rangle \approx 1.17 \times 10^{-12} E_R$, (c) $n_1 = 15,
    \langle{\Gamma^{sc}_{0,0}}\rangle \approx 1.11 \times 10^{-13} E_R$, and (d) $n_1 = 45, 
    \langle{\Gamma^{sc}_{0,0}}\rangle \approx 8.61 \times 10^{-13} E_R$. Left color bars represent the values for the quasi-free states, while the right ones -- the values for the dimer states.}
    \label{fig:Gamma_heatmap_band}
\end{figure}
\subsection{Numerical estimation of losses for two-particle dark state }
\label{subsection:2pBHint2}

The loss rate ${\Gamma}_{\textrm{tot}}$ is obtained from the FGR Eq.~\eqref{eq:gammafinal} by summation over all decay channels including different transverse modes. Among different possible final choices of the $\Lambda$-system channels, we find that if $(\tau_1,\tau_2)\in { (D,D),(D,B_+),(B_+,B+) }$ then no energy-resonant processes are allowed. The remaining possibilities are $(B_-,B_+),(B_-,B_-), (D,B_-)$, which allow for two-particle states with total energy close to 0. Here we focus on $(D,B_-)$, since we found that it gives the largest contribution to $\Gamma$, which is at least two orders of magnitude larger than the other two channels. 

In Section \ref{subsec:perpzero} we first discuss the rate $\Gamma^{sc}_{\bm\alpha_f,\bm\beta_f}$ for different final decay channels, for the lowest transverse modes $\bm\alpha_f=0, \bm\beta_f=0$. This provides a foundation for explaining the dependence of the final $\Gamma_{tot}$ on the system parameters, particularly $\tilde{\Omega}_c, \tilde{\Omega}_p.$

\begin{figure}[t]
    \includegraphics[width=0.99\linewidth]{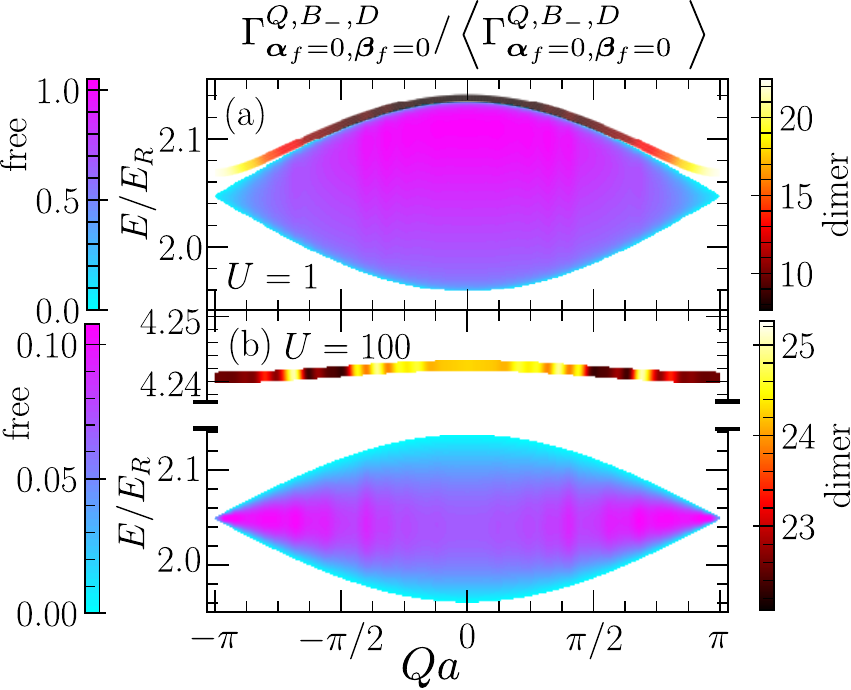}
    \caption{Dependence of a loss rate ${\Gamma}_{\bm\alpha_f=0,\bm\beta_f=0}^{Q,B_-,D}$ to a single channel with a lowest transverse state on the total quasimomentum $Q$ and energy $E$. The loss rate is normalized by its mean: ${\Gamma}_{\bm\alpha_f=0,\beta_f=0}^{Q,B_-,D}/\left \langle {\Gamma}_{\bm\alpha_f=0,\beta_f=0}^{Q,B_-,D} \right \rangle $. The laser parameters are assumed to give $\tilde{\Omega}_c = 4 \times 10^4 E_R$. The panels differ by the assumed interaction strength; (a) $U = 1$ ($g_0 \approx 0.0146$), $\langle{\Gamma}^{Q,B_,D}_{\bm\alpha_f=0,\bm\beta_f=0}\rangle \approx 1.81 \times 10^{-10} E_R$, and (b) $U = 100$ ($g_0 \approx 1.46$), $\langle{\Gamma}^{Q,B_,D}_{\bm\alpha_f=0,\bm\beta_f=0}\rangle \approx 1.85 \times 10^{-6} E_R$. Left colorbars represent the values for the quasi-free states, while the right ones -- the values for the dimer states.}
    \label{fig:Gamma_tot}
\end{figure}

\subsubsection{Loss rates for two-particle initial states}
\label{subsec:perpzero}

The decay of the initial state as in Eq.~\eqref{eqs:state} with $|\phi\rangle$ as in Eq.~\eqref{eqn:vqq_coeffs} occurs into an energy-resonant state for each set of parameters $n_1$, $n_2$, $\tau_1 = D$, and $\tau_2 = B_-$. To determine it, we use the spectral properties of the separate BO channels $D$ and $B_-$ -- see~\ref{sec:spchannels}. The energy spectrum of the $D$ channel starts at $1E_R$, and for $B_-$ the minimum is a large negative value given by Eq.~\ref{eq:3evelEplus}. For the values of $\tilde{\Omega}_c, \tilde{\Omega}_p$ considered here, $\tilde{\Omega}_c=O(10^4E_R), \tilde{\Omega}_p=O(10^3E_R)$ and $\Delta \ll \tilde{\Omega}_c$, the minimum of $B_-$ is (tens) of thousands of $E_R$ below zero. The $B_-$ has, for any quasimomentum, on the order of 100 eigenvalues with a negative energy. As a result, a two-particle non-interacting final state of near-zero energy can be realized with the $D$ channel eigenvalue index $n_1<n_1^\textrm{max}<\infty$. For our parameter regime, $n_1^\textrm{max}\approx 50 \divisionsymbol 150$. From the above, it follows that in order for the two-particle noninteracting $B_-,D$ state to have energy close to zero, either $D$ or $B_-$ component has to come from the part of the spectrum above the BO potential with a strong dependence on the $q_1$ or $q_2$ quasimomentum. \footnote{This also means that one of the two derivatives in Eqn.~\eqref{eqn:densityOfStates} dominates the other, avoiding the zero in the denominator}. Energy conservation selects one state from the continuum of states with a fixed $Q$ indexed by $q_1$ followed by $q_2=Q-q_1$.

\begin{figure}[t!]
    \centering
    \includegraphics[width=\linewidth]{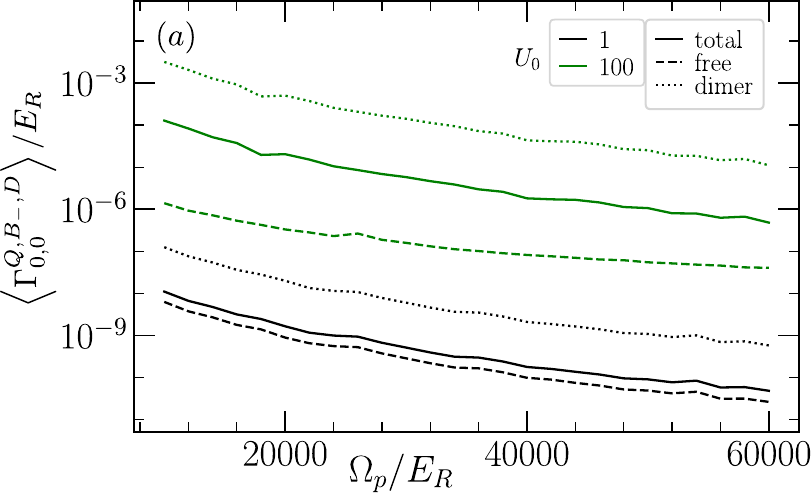}
    \includegraphics[width=\linewidth]{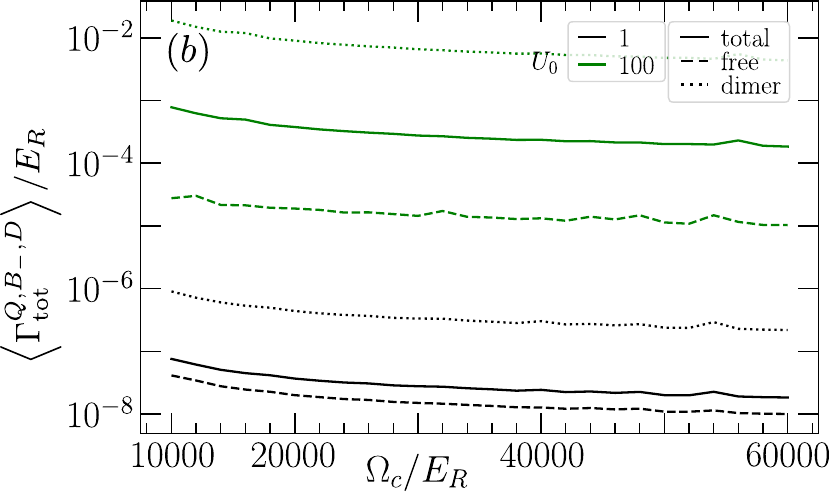}
    \caption{The dependence of total loss rates on $\tilde{\Omega}_c$ for $\tilde{\Omega}_p/\tilde{\Omega}_c=\epsilon=0.1$. The lattice size is $N=50$. Black and green lines correspond to, respectively, $U = 1$ ($g_0 \approx 0.0146$), and $U = 100$ ($g_0 \approx 1.46$). Solid, dashed, and dotted line denote averaging over, respectively, all initial states, quasi-free initial states and dimer initial states. Panel (a) shows the total loss rate for $\bm\alpha_f=\bm\beta_f=0$. The $\langle \Gamma_{0,0}^{Q,B_-,D}\rangle\sim\tilde{\Omega}_c^{-3.251\pm0.052}$ for $U/J=1$ and $\langle \Gamma_{0,0}^{Q,B_-,D}\rangle\sim\tilde{\Omega}_c^{-3.214\pm0.066}$  for $U/J=100$. Analogously quasi-free averages are $\langle \Gamma_{0,0}^{Q,B_-,D}\rangle_{qf}\sim\tilde{\Omega}_c^{-3.281\pm0.055}$ for $U/J=1$ and $\langle \Gamma_{0,0}^{Q,B_-,D}\rangle_{qf}\sim\tilde{\Omega}_c^{-2.001\pm0.041}$  for $U/J=100$. Panel (b) shows the total loss rate over all transverse modes (with $\hbar\omega_\perp=10E_R$). The $\langle \Gamma_{\textrm{tot}}^{Q,B_-,D}\rangle\sim\tilde{\Omega}_c^{-0.611\pm0.025}$ for $U/J=1$ and $\langle \Gamma_{\textrm{tot}}^{Q,B_-,D}\rangle\sim\tilde{\Omega}_c^{-0.610\pm0.027}$ for $U/J=100$. Analogously, quasi-free averages are $\langle \Gamma_{\textrm{tot}}^{Q,B_-,D}\rangle_{qf}\sim\tilde{\Omega}_c^{-0.623\pm0.018}$ for $U/J=1$ and $\langle \Gamma_{\textrm{tot}}^{Q,B_-,D}\rangle_{qf}\sim\tilde{\Omega}_c^{-0.459\pm0.054}$  for $U/J=100$. The exponents for the dimer averages are almost identical to the one of the total loss rate, both in (a) and (b).}
    \label{fig:Gamma_vs_Omega}
\end{figure}

In Fig.~\ref{fig:Gamma_heatmap_band} we show a color density plot of the loss rate $\Gamma^{sc}_{\bm\alpha_f=0,\bm\beta_f=0}$. In panels (a) through (d) we show the data for the final dark state eigenvalue index $n_1\in\{0,6,15,45\}$.
The particular points in each panel correspond to the corresponding energy $E$ and the total quasimomentum $Q=q_1+q_2$ of the initial state. The large lemon-shaped structure consists of $N(N-1)$ ``quasi-free'' states and is described by Eq.~\eqref{eq:E_free}. The small strip above is formed by $N$ ``dimer'' states; see Eq.~\eqref{eq:twoparticlestrip}. 

The values of $\Gamma^{sc}_{\bm\alpha_f=0,\bm\beta_f=0}$ have been independently normalized by the mean $\langle \Gamma^{sc}_{\bm\alpha_f=0,\bm\beta_f=0} \rangle $ specific to each panel.
The absolute values of $\Gamma^{sc}_{\bm\alpha_f=0,\bm\beta_f=0}$ indicated in figures for the parameters chosen are extremely small. 
This alone does not mean that the two-body losses are negligible, as here only a single decay channel is illustrated. The total $\Gamma_{\textrm{tot}}$ is a summation over all decay channels.
The dimer states are localized in the variable describing the relative position of the two particles. This makes the two-body interaction much more probable and on the order of $\approx N$ larger $\langle \Gamma^{sc}_{\bm\alpha_f=0,\bm\beta_f=0}\rangle$ than for the quasi-free states. However, it is important to note that this is offset by the smaller population of the dimer states compared to the quasi-free ones, which makes their contribution to the total decay rate mostly system size independent.

The $\langle \Gamma^{sc}_{\bm\alpha_f=0,\bm\beta_f=0}\rangle$ shows a significant variation for the quasi-free states. Also in the case of the dimer states the total variation is by tens of percent over the BZ. We have verified that for fixed decay channels, the variation of $\Gamma^{sc}_{\bm\alpha_f=0,\bm\beta_f=0}$ is mainly due to the matrix element $|\langle \psi_1,\psi_2 |\hat{V}_1|\phi \rangle |$. 

A similar effect has already been observed for single-particle losses for the operator $i \partial_x$ in \cite{Lacki2016}.
There, this translated to the quasimomentum-dependent lifetime of dark Bloch states $\psi_{n_i,q_i,D}$. 
In case of the contact interaction, the matrix element in the FGR is between two particle states where only the total quasi-momentum is conserved. The quasifree/dimer states are superpositions of all Bloch states with a given $Q=q_3+q_4$, see Eq.~\eqref{eqn:vqq_coeffs}. 
This makes the dependence of the matrix element more convoluted than in the single-particle case.
The dimer states, because they are partially localized, have significant overlap in many of the constituent Bloch states -- Eq.~\eqref{eqn:vqq_coeffs}. 
This means that any variation of the matrix element is effectively averaged over many $|\psi_{0,q_3,D},\psi_{0,q_4,D}\rangle$. 
In particular, in contrast to the case of the quasi-free initial states, for some values of $Q$, there exist $\Gamma^{sc}_{\bm\alpha_f,\bm\beta_f}$ that are non-negligible.

The strong dependence of the loss rate on $Q$ and the energy of the initial state persists when a partial sum is performed over the initial states. We sum over the final band index $n_1$ to get $\Gamma^{Q,B_-,D}_{\bm\alpha_f,\bm\beta_f}$.  
When $U/t=1$ the maximal loss rate within quasi-free states is for the states with $Qa \approx 0$ and the energy close to the top of the Bloch band. For strong interactions, the losses are maximized for $Qa \approx \pm \pi$. The strength of the interaction enters through the coefficients $v_{q_3q_4}$ in Eq.~\eqref{eqn:vqq_coeffs}. This makes the matrix element $|\langle \psi_1,\psi_2 |\hat{V}_1|\phi \rangle |$ interaction dependent beyond the trivial dependence via $\sim g_{3D}^2$ in Eq.~\eqref{eqn:matel}.
In Fig.~\ref{fig:Gamma_tot} we show it for two values of $U/t=1 \textrm{ and } 100$.

\subsection{Averaged loss rates}
\label{subsec:avglossrate}
We now discuss the loss rate averaged over energies and $Q$ of the initial state -- $\langle \Gamma_{0,0}^{Q,B_-,D} \rangle$. It is appropriate if experimental conditions do not allow for population of a single quantum state. The averaging is performed over $N(N-1)$ quasi-free states and $N$ dimer states leading to 
\begin{equation}
\langle \Gamma_{0,0}^{Q,B_-,D} \rangle = \frac{N-1}{N}\langle \Gamma_{0,0}^{Q,B_-,D} \rangle_{\textrm{qf}} + \frac{1}{N}\langle \Gamma_{0,0}^{Q,B_-,D} \rangle_{\textrm{dimer}}.
\label{eqn:Gammatotal0}
\end{equation}
The Fig.~\ref{fig:Gamma_vs_Omega} panel (a) shows the total loss rate for the lowest transverse mode $\langle \Gamma_{0,0}^{Q,B_-,D} \rangle$ together with the averages $\langle \Gamma_{0,0}^{Q,B_-,D} \rangle_{\textrm{dimer}}$ and $\langle \Gamma_{0,0}^{Q,B_-,D} \rangle_{\textrm{qf}}.$ All of them are shown for $U/t=1$ and $U/t=100$. 
When the decay channels are summed over final transverse modes $\bm\alpha_f, \bm\beta_f$ a similar equation holds:
\begin{equation}
\langle \Gamma_{\textrm{tot}}^{Q,B_-,D} \rangle = \frac{N-1}{N}\langle \Gamma_{\textrm{tot}}^{Q,B_-,D} \rangle_{\textrm{qf}} + \frac{1}{N}\langle \Gamma_{\textrm{tot}}^{Q,B_-,D} \rangle_{\textrm{dimer}}.
\label{eqn:Gammatotal}
\end{equation}
In the context set by the above equations we are able to make several observations. 

First, the loss rates $\langle \Gamma_{0,0}^{Q,B_-,D} \rangle $ and $\langle \Gamma_{0,0}^{Q,B_-,D} \rangle_{\textrm{qf}}$ are independent of $N$ which we have verified numerically. The $\langle \Gamma_{0,0}^{Q,B_-,D} \rangle_{\textrm{dimer}}$ increases proportionally to $N$, thus its contribution to $\langle \Gamma_{0,0}^{Q,B_-,D} \rangle$ remains $N$ independent due to the prefactor $\frac{1}{N}$ in Eq.~\eqref{eqn:Gammatotal0}. The same is true for the loss rates summed over the final transverse modes $\bm\alpha_f, \bm\beta_f$. This agrees with the reasoning given at the end of Section~\ref{subsec:fgrnumeric}.

Second, for a fixed $\tilde{\Omega}_p/\tilde{\Omega}_c = \epsilon= 0.1$  the $\langle \Gamma_{0,0}^{Q,B_-,D} \rangle \sim\tilde{\Omega}_c^{-3.251\pm0.052}$ for $U/J=1$ and $\sim\tilde{\Omega}_c^{-3.214\pm0.066}$ for $U/J=100$. When the summation over transverse modes is performed, the exponents are significantly reduced down to 
\begin{equation}
   \langle \Gamma_{\textrm{tot}}^{Q,B_-,D}\rangle\sim\tilde{\Omega}_c^{-0.611\pm0.025}, \quad U/J=1 
\end{equation}
and 
\begin{equation}
   \langle \Gamma_{\textrm{tot}}^{Q,B_-,D}\rangle\sim\tilde{\Omega}_c^{-0.610\pm0.027},\quad U/J=100.
\end{equation}
This is because as $\tilde{\Omega}_c$ and $\tilde{\Omega}_p$ increase, the summation of $\langle \Gamma_{\textrm{tot}}^{Q,B_-,D}\rangle$  as in Eq.~\eqref{eqn:transverseSummation} spans more and more decay channels. The summation is always finite: the maximal $n$ satisfies $n_{\textrm{max}}\hbar\omega_\perp \approx \min_x E_-(x)$ $ \approx -\sqrt{\Delta^2+4|\tilde{\Omega}_p|^2 +4|\Omega_c(x)|^2}/4$. The exact value of the exponents $-0.611\pm0.025$ and $-0.610\pm0.027$ depends only weakly on the transverse confinement. When $\hbar \omega_\perp = 100$ we have that it increases to
\begin{equation}
   \langle \Gamma_{\textrm{tot}}^{Q,B_-,D}\rangle\sim\tilde{\Omega}_c^{-0.890\pm0.030}, \quad U/J=1 
\end{equation}
and 
\begin{equation}
   \langle \Gamma_{\textrm{tot}}^{Q,B_-,D}\rangle\sim\tilde{\Omega}_c^{-0.886\pm0.039},\quad U/J=100.
\end{equation}
The losses also depend, for a fixed $\tilde{\Omega}_c,\tilde{\Omega}_p$, on $\omega_\perp$. We find that 
\begin{equation}
   \langle \Gamma_{\textrm{tot}}^{Q,B_-,D}\rangle\sim\omega_\perp^\gamma,\quad U/J=100,
\end{equation}
and the value of $\gamma$ is essentially equal $-1$ for $\Omega_p/\Omega_c=0.1$ and $\Omega_c\geq 2\cdot 10^4E_R$, for both $U/J=1$ and $U/J=100$

\begin{figure}
    \centering
    \includegraphics[width=8.4cm]{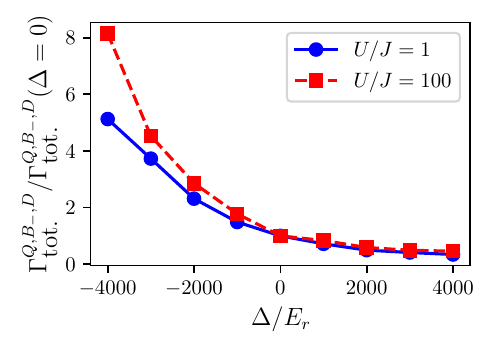}
    \caption{Panel a) The ratio of the decay rate $\Gamma^{Q,B_-,D}_{\textrm{tot}}$ for different $\Lambda$-system detuning $\Delta$ to the value for $\Delta=0$. The limiting value of $\Delta$ correspond to the value of $\Omega_p=4000E_r$, for $\epsilon=0.1$. For larger $|\Delta|$, one of the bright states approaches the dark state. For large negative $\Delta$, the loss rate per site is increased nearly by an order of magnitude.}
    \label{fig:plotDelta}
\end{figure}

As in \cite{Lacki2016}, the detuning $\Delta$ from the two-photon resonance in $\Lambda$ strongly affects the decay rate. For large negative $\Delta$, the $-2\Delta$ in Eq.~\eqref{eqn:3levels} is positive. As such, the bright-state channel $B_+$ moves to even higher positive energies. The threshold of a potential $E_-(x)$ in the $B_-$ channel originally at $\Omega_p/2$ changes to higher values, bringing it closer to the dark state manifold. This results in higher losses for a large negative $\Delta$, as shown in Fig.~\ref{fig:plotDelta}.

We have found numerically that if $\Delta$ is changed from $0$ to $4000E_R=\Omega_p$ for $\epsilon=0.1$, then the expected loss rate is approximately half for both $U/J=1$ and $U/J=100$. However, for $\Delta=-4000E_R$, the losses increase by a factor close to 5.1 for $U/J=1$ and to 8.1 for $U/J=100$.

\section{Conclusions and outlooks}
\label{sec:conclusions}
We have investigated two-particle losses due to contact interactions of bosons in the dark-state potential $V_{na}(x)$. The single particle realizations were based on strong lasers resulting in strong lasers forming the underlying $\Lambda$ system. The corresponding Rabi frequencies $\Omega_c(x)$ and $\Omega_p$ are on the order of tens of thousands of lattice recoil energy units. Single-particle losses in the $\Lambda$ system have been shown \cite{Lacki2016} to be reduced with an exponent of -2 (for a constant ratio of the two frequencies). In case of interaction-driven losses we observe a similar behavior that probability amplitudes are responsible for the dark state depletion, but we found that the value of the exponent is attenuated above $-1$. This is because the number of possible decay channels increases together with the amplitude of $\tilde{\Omega}_c$ and $\tilde{\Omega}_p$.  We also found that considering tighter harmonic confinement is a way to reduce the losses in our system; however, the relevant exponent is still close to $-1$, and therefore the loss rate shows a noticeable decrease with growing laser strength.
We have also found that detuning from the upper state in the $\Lambda$ system strongly affects the expected loss rate. The losses were strongly enhanced for the blue-retuned case and reduced by the opposite case. In general, changing the detuning from $-
\tilde{\Omega}_p$ to $
\tilde{\Omega}_p$ corresponds to a full order of magnitude difference in the loss rate.

A practical conclusion we draw from our calculation is that two-particle losses allow for the practical construction of bosonic systems in the dark-state potential for weak interactions. Specifically, this occurs when the Bose-Hubbard parameter ratio $U/J$ corresponds to a deep superfluid regime of the unit density phase diagram.  In that regime, we interpret the two-particle loss rate as the ``loss rate per site''. For much stronger interactions, our two-body loss rate $\left\langle\Gamma_{\textrm{tot}}^{Q,B_-,D}\right\rangle$ 
approaches values comparable or even larger than the single-particle loss rate reported in \cite{Lacki2016} [that is, $\Gamma = O(10^{-2}E_R)$]. This can be mitigated by increasing the values of
$\Omega_p, \Omega_c(x)$. For strong interaction, modeling many-body states via
a Gutzwiller-type product of the single-particle functions is not accurate.

Our calculations based on Fermi's Golden Rule can be more comprehensive by directly including many-body correlated interacting states. For example, in the unit filled Mott insulating phase, on the one hand, there is a strong interaction; on the other hand, the interaction prohibits double population of sites with variance of site population $\var n \sim 8J^2/U^2$. Another point that would benefit from additional analysis is the calculation of the loss rate based on multichannel scattering theory. The interaction-driven depopulation of two-particle dark Bloch states would manifest itself in a complex part of $g_{3D,0}$ in~Eq.~\eqref{eq:gtensordelta}. We plan to investigate these questions in further research.

\begin{acknowledgments}
    The authors appreciate the Polish high-performance computing infrastructure PLGrid (HPC Center: ACK Cyfronet AGH) for providing computer facilities and support within computational grant no. PLG/2023/016725 and PLG/2024/017241. We acknowledge support from
the National Science Center (Poland) via Opus Grant
No. 2019/35/B/ST2/00838
\end{acknowledgments}

%\bibliography{main}
\include{main.bbl}

\end{document}

%% file: main.bbl

%% file: main.bbl
\begin{thebibliography}{28}%
\makeatletter
\providecommand \@ifxundefined [1]{%
 \@ifx{#1\undefined}
}%
\providecommand \@ifnum [1]{%
 \ifnum #1\expandafter \@firstoftwo
 \else \expandafter \@secondoftwo
 \fi
}%
\providecommand \@ifx [1]{%
 \ifx #1\expandafter \@firstoftwo
 \else \expandafter \@secondoftwo
 \fi
}%
\providecommand \natexlab [1]{#1}%
\providecommand \enquote  [1]{``#1''}%
\providecommand \bibnamefont  [1]{#1}%
\providecommand \bibfnamefont [1]{#1}%
\providecommand \citenamefont [1]{#1}%
\providecommand \href@noop [0]{\@secondoftwo}%
\providecommand \href [0]{\begingroup \@sanitize@url \@href}%
\providecommand \@href[1]{\@@startlink{#1}\@@href}%
\providecommand \@@href[1]{\endgroup#1\@@endlink}%
\providecommand \@sanitize@url [0]{\catcode `\\12\catcode `\$12\catcode
  `\&12\catcode `\#12\catcode `\^12\catcode `\_12\catcode `\%12\relax}%
\providecommand \@@startlink[1]{}%
\providecommand \@@endlink[0]{}%
\providecommand \url  [0]{\begingroup\@sanitize@url \@url }%
\providecommand \@url [1]{\endgroup\@href {#1}{\urlprefix }}%
\providecommand \urlprefix  [0]{URL }%
\providecommand \Eprint [0]{\href }%
\providecommand \doibase [0]{https://doi.org/}%
\providecommand \selectlanguage [0]{\@gobble}%
\providecommand \bibinfo  [0]{\@secondoftwo}%
\providecommand \bibfield  [0]{\@secondoftwo}%
\providecommand \translation [1]{[#1]}%
\providecommand \BibitemOpen [0]{}%
\providecommand \bibitemStop [0]{}%
\providecommand \bibitemNoStop [0]{.\EOS\space}%
\providecommand \EOS [0]{\spacefactor3000\relax}%
\providecommand \BibitemShut  [1]{\csname bibitem#1\endcsname}%
\let\auto@bib@innerbib\@empty
%</preamble>
\bibitem [{\citenamefont {Greiner}\ \emph {et~al.}(2002)\citenamefont
  {Greiner}, \citenamefont {Mandel}, \citenamefont {Esslinger}, \citenamefont
  {H{\"a}nsch},\ and\ \citenamefont {Bloch}}]{Bloch2002}%
  \BibitemOpen
  \bibfield  {author} {\bibinfo {author} {\bibfnamefont {M.}~\bibnamefont
  {Greiner}}, \bibinfo {author} {\bibfnamefont {O.}~\bibnamefont {Mandel}},
  \bibinfo {author} {\bibfnamefont {T.}~\bibnamefont {Esslinger}}, \bibinfo
  {author} {\bibfnamefont {T.~W.}\ \bibnamefont {H{\"a}nsch}},\ and\ \bibinfo
  {author} {\bibfnamefont {I.}~\bibnamefont {Bloch}},\ }\bibfield  {title}
  {\bibinfo {title} {Quantum phase transition from a superfluid to a mott
  insulator in a gas of ultracold atoms},\ }\href@noop {} {\bibfield  {journal}
  {\bibinfo  {journal} {nature}\ }\textbf {\bibinfo {volume} {415}},\ \bibinfo
  {pages} {39} (\bibinfo {year} {2002})}\BibitemShut {NoStop}%
\bibitem [{\citenamefont {Bloch}\ \emph {et~al.}(2008)\citenamefont {Bloch},
  \citenamefont {Dalibard},\ and\ \citenamefont {Zwerger}}]{bloch2008many}%
  \BibitemOpen
  \bibfield  {author} {\bibinfo {author} {\bibfnamefont {I.}~\bibnamefont
  {Bloch}}, \bibinfo {author} {\bibfnamefont {J.}~\bibnamefont {Dalibard}},\
  and\ \bibinfo {author} {\bibfnamefont {W.}~\bibnamefont {Zwerger}},\
  }\bibfield  {title} {\bibinfo {title} {Many-body physics with ultracold
  gases},\ }\href@noop {} {\bibfield  {journal} {\bibinfo  {journal} {Reviews
  of modern physics}\ }\textbf {\bibinfo {volume} {80}},\ \bibinfo {pages}
  {885} (\bibinfo {year} {2008})}\BibitemShut {NoStop}%
\bibitem [{\citenamefont {Lewenstein}\ \emph {et~al.}(2012)\citenamefont
  {Lewenstein}, \citenamefont {Sanpera},\ and\ \citenamefont
  {Ahufinger}}]{lewenstein2012ultracold}%
  \BibitemOpen
  \bibfield  {author} {\bibinfo {author} {\bibfnamefont {M.}~\bibnamefont
  {Lewenstein}}, \bibinfo {author} {\bibfnamefont {A.}~\bibnamefont
  {Sanpera}},\ and\ \bibinfo {author} {\bibfnamefont {V.}~\bibnamefont
  {Ahufinger}},\ }\href@noop {} {\emph {\bibinfo {title} {Ultracold Atoms in
  Optical Lattices}}}\ (\bibinfo  {publisher} {OUP Oxford},\ \bibinfo {year}
  {2012})\BibitemShut {NoStop}%
\bibitem [{\citenamefont {Bloch}\ \emph {et~al.}(2012)\citenamefont {Bloch},
  \citenamefont {Dalibard},\ and\ \citenamefont
  {Nascimbene}}]{bloch2012quantum}%
  \BibitemOpen
  \bibfield  {author} {\bibinfo {author} {\bibfnamefont {I.}~\bibnamefont
  {Bloch}}, \bibinfo {author} {\bibfnamefont {J.}~\bibnamefont {Dalibard}},\
  and\ \bibinfo {author} {\bibfnamefont {S.}~\bibnamefont {Nascimbene}},\
  }\bibfield  {title} {\bibinfo {title} {Quantum simulations with ultracold
  quantum gases},\ }\href@noop {} {\bibfield  {journal} {\bibinfo  {journal}
  {Nature Physics}\ }\textbf {\bibinfo {volume} {8}},\ \bibinfo {pages} {267}
  (\bibinfo {year} {2012})}\BibitemShut {NoStop}%
\bibitem [{\citenamefont {Dutta}\ \emph {et~al.}(2015)\citenamefont {Dutta},
  \citenamefont {Gajda}, \citenamefont {Hauke}, \citenamefont {Lewenstein},
  \citenamefont {L{\"u}hmann}, \citenamefont {Malomed}, \citenamefont
  {Sowi{\'n}ski},\ and\ \citenamefont {Zakrzewski}}]{dutta2015}%
  \BibitemOpen
  \bibfield  {author} {\bibinfo {author} {\bibfnamefont {O.}~\bibnamefont
  {Dutta}}, \bibinfo {author} {\bibfnamefont {M.}~\bibnamefont {Gajda}},
  \bibinfo {author} {\bibfnamefont {P.}~\bibnamefont {Hauke}}, \bibinfo
  {author} {\bibfnamefont {M.}~\bibnamefont {Lewenstein}}, \bibinfo {author}
  {\bibfnamefont {D.-S.}\ \bibnamefont {L{\"u}hmann}}, \bibinfo {author}
  {\bibfnamefont {B.~A.}\ \bibnamefont {Malomed}}, \bibinfo {author}
  {\bibfnamefont {T.}~\bibnamefont {Sowi{\'n}ski}},\ and\ \bibinfo {author}
  {\bibfnamefont {J.}~\bibnamefont {Zakrzewski}},\ }\bibfield  {title}
  {\bibinfo {title} {Non-standard hubbard models in optical lattices: a
  review},\ }\href@noop {} {\bibfield  {journal} {\bibinfo  {journal} {Reports
  on Progress in Physics}\ }\textbf {\bibinfo {volume} {78}},\ \bibinfo {pages}
  {066001} (\bibinfo {year} {2015})}\BibitemShut {NoStop}%
\bibitem [{\citenamefont {Fedichev}\ \emph {et~al.}(1996)\citenamefont
  {Fedichev}, \citenamefont {Kagan}, \citenamefont {Shlyapnikov},\ and\
  \citenamefont {Walraven}}]{fedichev1996influence}%
  \BibitemOpen
  \bibfield  {author} {\bibinfo {author} {\bibfnamefont {P.}~\bibnamefont
  {Fedichev}}, \bibinfo {author} {\bibfnamefont {Y.}~\bibnamefont {Kagan}},
  \bibinfo {author} {\bibfnamefont {G.}~\bibnamefont {Shlyapnikov}},\ and\
  \bibinfo {author} {\bibfnamefont {J.}~\bibnamefont {Walraven}},\ }\bibfield
  {title} {\bibinfo {title} {Influence of nearly resonant light on the
  scattering length in low-temperature atomic gases},\ }\href@noop {}
  {\bibfield  {journal} {\bibinfo  {journal} {Physical review letters}\
  }\textbf {\bibinfo {volume} {77}},\ \bibinfo {pages} {2913} (\bibinfo {year}
  {1996})}\BibitemShut {NoStop}%
\bibitem [{\citenamefont {Bohn}\ and\ \citenamefont
  {Julienne}(1997)}]{bohn1997prospects}%
  \BibitemOpen
  \bibfield  {author} {\bibinfo {author} {\bibfnamefont {J.~L.}\ \bibnamefont
  {Bohn}}\ and\ \bibinfo {author} {\bibfnamefont {P.~S.}\ \bibnamefont
  {Julienne}},\ }\bibfield  {title} {\bibinfo {title} {Prospects for
  influencing scattering lengths with far-off-resonant light},\ }\href@noop {}
  {\bibfield  {journal} {\bibinfo  {journal} {Physical Review A}\ }\textbf
  {\bibinfo {volume} {56}},\ \bibinfo {pages} {1486} (\bibinfo {year}
  {1997})}\BibitemShut {NoStop}%
\bibitem [{\citenamefont {Theis}\ \emph {et~al.}(2004)\citenamefont {Theis},
  \citenamefont {Thalhammer}, \citenamefont {Winkler}, \citenamefont {Hellwig},
  \citenamefont {Ruff}, \citenamefont {Grimm},\ and\ \citenamefont
  {Denschlag}}]{theis2004tuning}%
  \BibitemOpen
  \bibfield  {author} {\bibinfo {author} {\bibfnamefont {M.}~\bibnamefont
  {Theis}}, \bibinfo {author} {\bibfnamefont {G.}~\bibnamefont {Thalhammer}},
  \bibinfo {author} {\bibfnamefont {K.}~\bibnamefont {Winkler}}, \bibinfo
  {author} {\bibfnamefont {M.}~\bibnamefont {Hellwig}}, \bibinfo {author}
  {\bibfnamefont {G.}~\bibnamefont {Ruff}}, \bibinfo {author} {\bibfnamefont
  {R.}~\bibnamefont {Grimm}},\ and\ \bibinfo {author} {\bibfnamefont {J.~H.}\
  \bibnamefont {Denschlag}},\ }\bibfield  {title} {\bibinfo {title} {Tuning the
  scattering length with an optically induced feshbach resonance},\ }\href@noop
  {} {\bibfield  {journal} {\bibinfo  {journal} {Physical Review Letters}\
  }\textbf {\bibinfo {volume} {93}},\ \bibinfo {pages} {123001} (\bibinfo
  {year} {2004})}\BibitemShut {NoStop}%
\bibitem [{\citenamefont {Inouye}\ \emph {et~al.}(1998)\citenamefont {Inouye},
  \citenamefont {Andrews}, \citenamefont {Stenger}, \citenamefont {Miesner},
  \citenamefont {Stamper-Kurn},\ and\ \citenamefont
  {Ketterle}}]{inouye1998observation}%
  \BibitemOpen
  \bibfield  {author} {\bibinfo {author} {\bibfnamefont {S.}~\bibnamefont
  {Inouye}}, \bibinfo {author} {\bibfnamefont {M.}~\bibnamefont {Andrews}},
  \bibinfo {author} {\bibfnamefont {J.}~\bibnamefont {Stenger}}, \bibinfo
  {author} {\bibfnamefont {H.-J.}\ \bibnamefont {Miesner}}, \bibinfo {author}
  {\bibfnamefont {D.~M.}\ \bibnamefont {Stamper-Kurn}},\ and\ \bibinfo {author}
  {\bibfnamefont {W.}~\bibnamefont {Ketterle}},\ }\bibfield  {title} {\bibinfo
  {title} {Observation of feshbach resonances in a bose--einstein condensate},\
  }\href@noop {} {\bibfield  {journal} {\bibinfo  {journal} {Nature}\ }\textbf
  {\bibinfo {volume} {392}},\ \bibinfo {pages} {151} (\bibinfo {year}
  {1998})}\BibitemShut {NoStop}%
\bibitem [{\citenamefont {Chin}\ \emph {et~al.}(2010)\citenamefont {Chin},
  \citenamefont {Grimm}, \citenamefont {Julienne},\ and\ \citenamefont
  {Tiesinga}}]{chin2010feshbach}%
  \BibitemOpen
  \bibfield  {author} {\bibinfo {author} {\bibfnamefont {C.}~\bibnamefont
  {Chin}}, \bibinfo {author} {\bibfnamefont {R.}~\bibnamefont {Grimm}},
  \bibinfo {author} {\bibfnamefont {P.}~\bibnamefont {Julienne}},\ and\
  \bibinfo {author} {\bibfnamefont {E.}~\bibnamefont {Tiesinga}},\ }\bibfield
  {title} {\bibinfo {title} {Feshbach resonances in ultracold gases},\
  }\href@noop {} {\bibfield  {journal} {\bibinfo  {journal} {Reviews of Modern
  Physics}\ }\textbf {\bibinfo {volume} {82}},\ \bibinfo {pages} {1225}
  (\bibinfo {year} {2010})}\BibitemShut {NoStop}%
\bibitem [{\citenamefont {Ciury{\l}o}\ \emph {et~al.}(2005)\citenamefont
  {Ciury{\l}o}, \citenamefont {Tiesinga},\ and\ \citenamefont
  {Julienne}}]{ciurylo2005optical}%
  \BibitemOpen
  \bibfield  {author} {\bibinfo {author} {\bibfnamefont {R.}~\bibnamefont
  {Ciury{\l}o}}, \bibinfo {author} {\bibfnamefont {E.}~\bibnamefont
  {Tiesinga}},\ and\ \bibinfo {author} {\bibfnamefont {P.}~\bibnamefont
  {Julienne}},\ }\bibfield  {title} {\bibinfo {title} {Optical tuning of the
  scattering length of cold alkaline-earth-metal atoms},\ }\href@noop {}
  {\bibfield  {journal} {\bibinfo  {journal} {Physical Review A—Atomic,
  Molecular, and Optical Physics}\ }\textbf {\bibinfo {volume} {71}},\ \bibinfo
  {pages} {030701} (\bibinfo {year} {2005})}\BibitemShut {NoStop}%
\bibitem [{\citenamefont {Jaksch}\ \emph
  {et~al.}(1998{\natexlab{a}})\citenamefont {Jaksch}, \citenamefont {Bruder},
  \citenamefont {Cirac}, \citenamefont {Gardiner},\ and\ \citenamefont
  {Zoller}}]{jaksch1998cold}%
  \BibitemOpen
  \bibfield  {author} {\bibinfo {author} {\bibfnamefont {D.}~\bibnamefont
  {Jaksch}}, \bibinfo {author} {\bibfnamefont {C.}~\bibnamefont {Bruder}},
  \bibinfo {author} {\bibfnamefont {J.~I.}\ \bibnamefont {Cirac}}, \bibinfo
  {author} {\bibfnamefont {C.~W.}\ \bibnamefont {Gardiner}},\ and\ \bibinfo
  {author} {\bibfnamefont {P.}~\bibnamefont {Zoller}},\ }\bibfield  {title}
  {\bibinfo {title} {Cold bosonic atoms in optical lattices},\ }\href@noop {}
  {\bibfield  {journal} {\bibinfo  {journal} {Physical Review Letters}\
  }\textbf {\bibinfo {volume} {81}},\ \bibinfo {pages} {3108} (\bibinfo {year}
  {1998}{\natexlab{a}})}\BibitemShut {NoStop}%
\bibitem [{\citenamefont {M{\"u}ller}\ \emph {et~al.}(2007)\citenamefont
  {M{\"u}ller}, \citenamefont {F{\"o}lling}, \citenamefont {Widera},\ and\
  \citenamefont {Bloch}}]{muller2007state}%
  \BibitemOpen
  \bibfield  {author} {\bibinfo {author} {\bibfnamefont {T.}~\bibnamefont
  {M{\"u}ller}}, \bibinfo {author} {\bibfnamefont {S.}~\bibnamefont
  {F{\"o}lling}}, \bibinfo {author} {\bibfnamefont {A.}~\bibnamefont
  {Widera}},\ and\ \bibinfo {author} {\bibfnamefont {I.}~\bibnamefont
  {Bloch}},\ }\bibfield  {title} {\bibinfo {title} {State preparation and
  dynamics of ultracold atoms in higher lattice orbitals},\ }\href@noop {}
  {\bibfield  {journal} {\bibinfo  {journal} {Physical Review Letters}\
  }\textbf {\bibinfo {volume} {99}},\ \bibinfo {pages} {200405} (\bibinfo
  {year} {2007})}\BibitemShut {NoStop}%
\bibitem [{\citenamefont {Johnson}\ \emph {et~al.}(2009)\citenamefont
  {Johnson}, \citenamefont {Tiesinga}, \citenamefont {Porto},\ and\
  \citenamefont {Williams}}]{johnson2009effective}%
  \BibitemOpen
  \bibfield  {author} {\bibinfo {author} {\bibfnamefont {P.}~\bibnamefont
  {Johnson}}, \bibinfo {author} {\bibfnamefont {E.}~\bibnamefont {Tiesinga}},
  \bibinfo {author} {\bibfnamefont {J.~V.}\ \bibnamefont {Porto}},\ and\
  \bibinfo {author} {\bibfnamefont {C.~J.}\ \bibnamefont {Williams}},\
  }\bibfield  {title} {\bibinfo {title} {Effective three-body interactions of
  neutral bosons in optical lattices},\ }\href@noop {} {\bibfield  {journal}
  {\bibinfo  {journal} {New Journal of Physics}\ }\textbf {\bibinfo {volume}
  {11}},\ \bibinfo {pages} {093022} (\bibinfo {year} {2009})}\BibitemShut
  {NoStop}%
\bibitem [{\citenamefont {Mering}\ and\ \citenamefont
  {Fleischhauer}(2011)}]{mering2011multiband}%
  \BibitemOpen
  \bibfield  {author} {\bibinfo {author} {\bibfnamefont {A.}~\bibnamefont
  {Mering}}\ and\ \bibinfo {author} {\bibfnamefont {M.}~\bibnamefont
  {Fleischhauer}},\ }\bibfield  {title} {\bibinfo {title} {Multiband and
  nonlinear hopping corrections to the three-dimensional bose-fermi-hubbard
  model},\ }\href@noop {} {\bibfield  {journal} {\bibinfo  {journal} {Physical
  Review A—Atomic, Molecular, and Optical Physics}\ }\textbf {\bibinfo
  {volume} {83}},\ \bibinfo {pages} {063630} (\bibinfo {year}
  {2011})}\BibitemShut {NoStop}%
\bibitem [{\citenamefont {L{\"u}hmann}\ \emph {et~al.}(2012)\citenamefont
  {L{\"u}hmann}, \citenamefont {J{\"u}rgensen},\ and\ \citenamefont
  {Sengstock}}]{luhmann2012multi}%
  \BibitemOpen
  \bibfield  {author} {\bibinfo {author} {\bibfnamefont {D.-S.}\ \bibnamefont
  {L{\"u}hmann}}, \bibinfo {author} {\bibfnamefont {O.}~\bibnamefont
  {J{\"u}rgensen}},\ and\ \bibinfo {author} {\bibfnamefont {K.}~\bibnamefont
  {Sengstock}},\ }\bibfield  {title} {\bibinfo {title} {Multi-orbital and
  density-induced tunneling of bosons in optical lattices},\ }\href@noop {}
  {\bibfield  {journal} {\bibinfo  {journal} {New Journal of Physics}\ }\textbf
  {\bibinfo {volume} {14}},\ \bibinfo {pages} {033021} (\bibinfo {year}
  {2012})}\BibitemShut {NoStop}%
\bibitem [{\citenamefont {Isacsson}\ and\ \citenamefont
  {Girvin}(2005)}]{isacsson2005multiflavor}%
  \BibitemOpen
  \bibfield  {author} {\bibinfo {author} {\bibfnamefont {A.}~\bibnamefont
  {Isacsson}}\ and\ \bibinfo {author} {\bibfnamefont {S.}~\bibnamefont
  {Girvin}},\ }\bibfield  {title} {\bibinfo {title} {Multiflavor bosonic
  hubbard models in the first excited bloch band of an optical lattice},\
  }\href@noop {} {\bibfield  {journal} {\bibinfo  {journal} {Physical Review
  A}\ }\textbf {\bibinfo {volume} {72}},\ \bibinfo {pages} {053604} (\bibinfo
  {year} {2005})}\BibitemShut {NoStop}%
\bibitem [{\citenamefont {Wirth}\ \emph {et~al.}(2011)\citenamefont {Wirth},
  \citenamefont {{\"O}lschl{\"a}ger},\ and\ \citenamefont
  {Hemmerich}}]{wirth2011evidence}%
  \BibitemOpen
  \bibfield  {author} {\bibinfo {author} {\bibfnamefont {G.}~\bibnamefont
  {Wirth}}, \bibinfo {author} {\bibfnamefont {M.}~\bibnamefont
  {{\"O}lschl{\"a}ger}},\ and\ \bibinfo {author} {\bibfnamefont
  {A.}~\bibnamefont {Hemmerich}},\ }\bibfield  {title} {\bibinfo {title}
  {Evidence for orbital superfluidity in the p-band of a bipartite optical
  square lattice},\ }\href@noop {} {\bibfield  {journal} {\bibinfo  {journal}
  {Nature Physics}\ }\textbf {\bibinfo {volume} {7}},\ \bibinfo {pages} {147}
  (\bibinfo {year} {2011})}\BibitemShut {NoStop}%
\bibitem [{\citenamefont {Li}\ and\ \citenamefont {Liu}(2016)}]{li2016physics}%
  \BibitemOpen
  \bibfield  {author} {\bibinfo {author} {\bibfnamefont {X.}~\bibnamefont
  {Li}}\ and\ \bibinfo {author} {\bibfnamefont {W.~V.}\ \bibnamefont {Liu}},\
  }\bibfield  {title} {\bibinfo {title} {Physics of higher orbital bands in
  optical lattices: a review},\ }\href@noop {} {\bibfield  {journal} {\bibinfo
  {journal} {Reports on Progress in Physics}\ }\textbf {\bibinfo {volume}
  {79}},\ \bibinfo {pages} {116401} (\bibinfo {year} {2016})}\BibitemShut
  {NoStop}%
\bibitem [{\citenamefont {{\L}{\k{a}}cki}\ \emph {et~al.}(2016)\citenamefont
  {{\L}{\k{a}}cki}, \citenamefont {Baranov}, \citenamefont {Pichler},\ and\
  \citenamefont {Zoller}}]{Lacki2016}%
  \BibitemOpen
  \bibfield  {author} {\bibinfo {author} {\bibfnamefont {M.}~\bibnamefont
  {{\L}{\k{a}}cki}}, \bibinfo {author} {\bibfnamefont {M.}~\bibnamefont
  {Baranov}}, \bibinfo {author} {\bibfnamefont {H.}~\bibnamefont {Pichler}},\
  and\ \bibinfo {author} {\bibfnamefont {P.}~\bibnamefont {Zoller}},\
  }\bibfield  {title} {\bibinfo {title} {Nanoscale 'dark state' optical
  potentials for cold atoms},\ }\href@noop {} {\bibfield  {journal} {\bibinfo
  {journal} {Physical review letters}\ }\textbf {\bibinfo {volume} {117}},\
  \bibinfo {pages} {233001} (\bibinfo {year} {2016})}\BibitemShut {NoStop}%
\bibitem [{\citenamefont {Dalibard}\ \emph {et~al.}(2011)\citenamefont
  {Dalibard}, \citenamefont {Gerbier}, \citenamefont {Juzeli{\=u}nas},\ and\
  \citenamefont {{\"O}hberg}}]{dalibard2011colloquium}%
  \BibitemOpen
  \bibfield  {author} {\bibinfo {author} {\bibfnamefont {J.}~\bibnamefont
  {Dalibard}}, \bibinfo {author} {\bibfnamefont {F.}~\bibnamefont {Gerbier}},
  \bibinfo {author} {\bibfnamefont {G.}~\bibnamefont {Juzeli{\=u}nas}},\ and\
  \bibinfo {author} {\bibfnamefont {P.}~\bibnamefont {{\"O}hberg}},\ }\bibfield
   {title} {\bibinfo {title} {Colloquium: Artificial gauge potentials for
  neutral atoms},\ }\href@noop {} {\bibfield  {journal} {\bibinfo  {journal}
  {Reviews of Modern Physics}\ }\textbf {\bibinfo {volume} {83}},\ \bibinfo
  {pages} {1523} (\bibinfo {year} {2011})}\BibitemShut {NoStop}%
\bibitem [{\citenamefont {Yang}\ \emph {et~al.}(2018)\citenamefont {Yang},
  \citenamefont {Vasilyev}, \citenamefont {Laflamme}, \citenamefont {Baranov},\
  and\ \citenamefont {Zoller}}]{Yang2018}%
  \BibitemOpen
  \bibfield  {author} {\bibinfo {author} {\bibfnamefont {D.}~\bibnamefont
  {Yang}}, \bibinfo {author} {\bibfnamefont {D.~V.}\ \bibnamefont {Vasilyev}},
  \bibinfo {author} {\bibfnamefont {C.}~\bibnamefont {Laflamme}}, \bibinfo
  {author} {\bibfnamefont {M.~A.}\ \bibnamefont {Baranov}},\ and\ \bibinfo
  {author} {\bibfnamefont {P.}~\bibnamefont {Zoller}},\ }\bibfield  {title}
  {\bibinfo {title} {Quantum scanning microscope for cold atoms},\ }\href@noop
  {} {\bibfield  {journal} {\bibinfo  {journal} {Physical Review A}\ }\textbf
  {\bibinfo {volume} {98}},\ \bibinfo {pages} {023852} (\bibinfo {year}
  {2018})}\BibitemShut {NoStop}%
\bibitem [{\citenamefont {Valiente}\ and\ \citenamefont
  {Petrosyan}(2008{\natexlab{a}})}]{praca2008}%
  \BibitemOpen
  \bibfield  {author} {\bibinfo {author} {\bibfnamefont {M.}~\bibnamefont
  {Valiente}}\ and\ \bibinfo {author} {\bibfnamefont {D.}~\bibnamefont
  {Petrosyan}},\ }\bibfield  {title} {\bibinfo {title} {Two-particle states in
  the hubbard model},\ }\href@noop {} {\bibfield  {journal} {\bibinfo
  {journal} {Journal of Physics B: Atomic, Molecular and Optical Physics}\
  }\textbf {\bibinfo {volume} {41}},\ \bibinfo {pages} {161002} (\bibinfo
  {year} {2008}{\natexlab{a}})}\BibitemShut {NoStop}%
\bibitem [{\citenamefont {Kohn}(1959)}]{Kohn1959}%
  \BibitemOpen
  \bibfield  {author} {\bibinfo {author} {\bibfnamefont {W.}~\bibnamefont
  {Kohn}},\ }\bibfield  {title} {\bibinfo {title} {Analytic properties of bloch
  waves and wannier functions},\ }\href@noop {} {\bibfield  {journal} {\bibinfo
   {journal} {Physical Review}\ }\textbf {\bibinfo {volume} {115}},\ \bibinfo
  {pages} {809} (\bibinfo {year} {1959})}\BibitemShut {NoStop}%
\bibitem [{\citenamefont {Jaksch}\ and\ \citenamefont
  {Zoller}(2005)}]{Jaksch2005}%
  \BibitemOpen
  \bibfield  {author} {\bibinfo {author} {\bibfnamefont {D.}~\bibnamefont
  {Jaksch}}\ and\ \bibinfo {author} {\bibfnamefont {P.}~\bibnamefont
  {Zoller}},\ }\bibfield  {title} {\bibinfo {title} {The cold atom hubbard
  toolbox},\ }\href@noop {} {\bibfield  {journal} {\bibinfo  {journal} {Annals
  of physics}\ }\textbf {\bibinfo {volume} {315}},\ \bibinfo {pages} {52}
  (\bibinfo {year} {2005})}\BibitemShut {NoStop}%
\bibitem [{\citenamefont {Jaksch}\ \emph
  {et~al.}(1998{\natexlab{b}})\citenamefont {Jaksch}, \citenamefont {Bruder},
  \citenamefont {Cirac}, \citenamefont {Gardiner},\ and\ \citenamefont
  {Zoller}}]{Jaksch1998}%
  \BibitemOpen
  \bibfield  {author} {\bibinfo {author} {\bibfnamefont {D.}~\bibnamefont
  {Jaksch}}, \bibinfo {author} {\bibfnamefont {C.}~\bibnamefont {Bruder}},
  \bibinfo {author} {\bibfnamefont {J.~I.}\ \bibnamefont {Cirac}}, \bibinfo
  {author} {\bibfnamefont {C.~W.}\ \bibnamefont {Gardiner}},\ and\ \bibinfo
  {author} {\bibfnamefont {P.}~\bibnamefont {Zoller}},\ }\bibfield  {title}
  {\bibinfo {title} {Cold bosonic atoms in optical lattices},\ }\href@noop {}
  {\bibfield  {journal} {\bibinfo  {journal} {Physical Review Letters}\
  }\textbf {\bibinfo {volume} {81}},\ \bibinfo {pages} {3108} (\bibinfo {year}
  {1998}{\natexlab{b}})}\BibitemShut {NoStop}%
\bibitem [{\citenamefont {Valiente}\ and\ \citenamefont
  {Petrosyan}(2008{\natexlab{b}})}]{Valiente2008}%
  \BibitemOpen
  \bibfield  {author} {\bibinfo {author} {\bibfnamefont {M.}~\bibnamefont
  {Valiente}}\ and\ \bibinfo {author} {\bibfnamefont {D.}~\bibnamefont
  {Petrosyan}},\ }\bibfield  {title} {\bibinfo {title} {Two-particle states in
  the hubbard model},\ }\href@noop {} {\bibfield  {journal} {\bibinfo
  {journal} {Journal of Physics B: Atomic, Molecular and Optical Physics}\
  }\textbf {\bibinfo {volume} {41}},\ \bibinfo {pages} {161002} (\bibinfo
  {year} {2008}{\natexlab{b}})}\BibitemShut {NoStop}%
\bibitem [{Note1()}]{Note1}%
  \BibitemOpen
  \bibinfo {note} {This also means that one of the two derivatives in
  Eqn.~\protect \eqref {eqn:densityOfStates} dominates the other, avoiding the
  zero in the denominator}\BibitemShut {NoStop}%
\end{thebibliography}
